\documentclass[preprint,aip]{revtex4-1}
\usepackage{amsmath}
\usepackage{amssymb}
\usepackage{graphicx,color}
\renewcommand{\theequation}{\arabic{section}.\arabic{equation}}

\begin{document}
\title{Impact of environmentally induced fluctuations on quantum mechanically mixed electronic and vibrational pigment states in photosynthetic energy transfer and 2D electronic spectra}

\author{Yuta Fujihashi}
\affiliation{Institute for Molecular Science, National Institutes of Natural Sciences, Okazaki 444-8585, Japan}

\author{Graham R. Fleming}
\affiliation{Department of Chemistry, University of California, Berkeley and Physical Biosciences Division, Lawrence Berkeley National Laboratory, Berkeley, CA 94720, USA}

\author{Akihito Ishizaki}
\email{ishizaki@ims.ac.jp}
\affiliation{Institute for Molecular Science, National Institutes of Natural Sciences, Okazaki 444-8585, Japan}

\begin{abstract}
Recently, nuclear vibrational contribution signatures in 2D electronic spectroscopy have attracted considerable interest, in particular as regards interpretation of the oscillatory transients observed in light-harvesting complexes. These transients have dephasing times that persist for much longer than theoretically predicted electronic coherence lifetime. As a plausible explanation for this long-lived spectral beating in 2D electronic spectra, quantum-mechanically mixed electronic and vibrational states (vibronic excitons) were proposed by Christensson {\it et al.} [J. Phys. Chem. B {\bf 116}, 7449 (2012)] and have since been explored. In this work, we address a dimer which produces little beating of electronic origin in the absence of vibronic contributions, and examine the impact of protein-induced fluctuations upon electronic-vibrational quantum mixtures by calculating the electronic energy transfer dynamics and 2D electronic spectra in a numerically accurate manner. It is found that, at cryogenic temperatures, the electronic-vibrational quantum mixtures are rather robust, even under the influence of the fluctuations and despite the small Huang-Rhys factors of the Franck-Condon active vibrational modes. This results in long-lasting beating behavior of vibrational origin in the 2D electronic spectra. At physiological temperatures, however, the fluctuations eradicate the mixing and, hence, the beating in the 2D spectra disappears. Further, it is demonstrated that such electronic-vibrational quantum mixtures do not necessarily play a significant role in electronic energy transfer dynamics, despite contributing to the enhancement of long-lived quantum beating in 2D electronic spectra, contrary to speculations in recent publications.
\end{abstract}
\maketitle
\section{Introduction}
\setcounter{equation}{0}

Recent observations of long-lived beating phenomena in the two-dimensional (2D) electronic spectra \cite{Engel:2007hb, Calhoun:2009bn, Collini:2010fy, Panitchayangkoon:2010fw,Panitchayangkoon:2011cs, SchlauCohen:2012dn, Westenhoff:2012fi,Dawlaty:2012fs,Fuller:2014iz,Romero:2014jm} of photosynthetic pigment-protein complexes have stimulated a considerable increase in activity in the interdisciplinary field of molecular science and quantum physics. \cite{Cheng:2009co,Ishizaki:2010fx, Ishizaki:2012kf, Jang:2012fi, Huelga:2013eu, Chenu:2015gy} It has generally been assumed that electronic coherence decays sufficiently rapidly that it does not affect the nature of electronic energy transfer in photosynthetic light harvesting systems. Nevertheless, these recent experiments have attributed long-lived quantum beats to an electronic origin. Initially, 2D electronic spectroscopic experiments were conducted for Fenna-Mathews-Olson (FMO) pigment-protein complexes isolated from green sulfur bacteria at a cryogenic temperature, 77\,K, \cite{Brixner:2005wu,Engel:2007hb} and revealed the presence of quantum beats persisting for a minimum of 660\,fs. \cite{Engel:2007hb} However, it is generally thought that coherence at physiological temperatures is fragile compared to that at cryogenic temperatures, because environmental fluctuation amplitudes increase with increasing temperature. To clarify this issue, electronic coherence lifetimes in the FMO complex were theoretically examined, and it was predicted that electronic coherence in this complex could persist for 700 and 300\,fs at 77 and 300\,K, respectively. \cite{Ishizaki:2009ky} 
These theoretical predictions were consistent with newer experimental results at physiological temperatures for the FMO complex. \cite{Panitchayangkoon:2010fw} 
However, Panitchayangkoon {\it et al.}\cite{Panitchayangkoon:2010fw} demonstrated that quantum beats in this complex persist for 1.5 ps at cryogenic temperatures, although the theoretical model does not produce electronic coherence with this lifetime. Hence, nuclear vibrational contribution signatures in 2D spectroscopy have attracted considerable interest of late, in particular as regards interpretations of the oscillatory transients observed in light-harvesting complexes, which persist for significantly longer than the predicted electronic dephasing times. Several independent publications have alluded to nuclear vibrational effects as an explanation for this long-lived beating behavior. \cite{Turner:2011ef,Christensson:2011ht, Christensson:2012gp, Kolli:2012ip,YuenZhou:2012hu, CaycedoSoler:2012ib, Butkus:2012hn, Butkus:2013fy, Tiwari:2013dt, Chenu:2013cf, Kreisbeck:2013jva, Chin:2013ia, Tiwari:2013dt, Plenio:2013bg,Rivera:2013kb,Seibt:2013dp,Halpin:2014jd, Tempelaar:2014vu,OReilly:2014it,Perlik:2014bd,Dijkstra:2015er,Nalbach:2015ii,Schroter:2015je}

In photosynthetic pigment-protein complexes, the Huang-Rhys factors of chlorophyll (Chl) and bacteriochlorophyll (BChl) are generally thought to be small, \cite{Wendling:2000ha,Ratsep:2007fq,Ratsep:2011cq,Schulze:2014iv} suggesting that the photophysics therein is primarily electronic in nature, rather than vibrational. Indeed, the recent 2D electronic experiments on BChl molecules in solution did not find significant vibrational coherence.\cite{Fransted:2012fs} In relation to this, Christensson {\it et al.} \cite{Christensson:2012gp} proposed that resonance between electronic states and active Franck-Condon vibrational states serves to create vibronic excitons, i.e., quantum mechanically mixed electronic and vibrational states. Such states have vibrational characteristics and enhanced transition dipole moments owing to intensity borrowing from the strong electronic transitions. Hence, it was suggested that coherent excitation of the vibronic excitons produces oscillations in the 2D signal that exhibit picosecond dephasing times. Tiwari, Peters, and Jonas \cite{Tiwari:2013dt} commented that the excitonically mixed electronic and vibrational states also lead to an enhancement of the vibrational coherence excitation in the electronic ground state. It was also argued that this effect could explain the long-lived oscillations in the FMO complex. Regarding electronic energy transfer dynamics, Womick and Moran \cite{Womick:2011iw} demonstrated that a vibrational-electronic resonance enhances the energy transfer rate in a cyanobacterial light-harvesting protein, allophycocyanin, with the use of the vibronic exciton model.

Quantum mechanically mixed electronic and vibrational states, or vibronic excitons, are plausible as an explanation for long-lived spectral beating in 2D electronic spectra. However, a question naturally arises concerning the interplay between the electronic-vibrational resonance and electronic energy fluctuations induced by the environmental dynamics. \cite{Ishizaki:2010ft} In general, the energy eigenstates due to the quantum mixing of electronic and vibronic excitations are obtained via diagonalization of the Hamiltonian comprising the Franck-Condon transition energies and electronic interactions. It should be noted that these are independent of any environmental factors such as temperature, reorganization energy, and fluctuations. Ishizaki and Fleming \cite{Ishizaki:2010ft, Ishizaki:2012kf} characterized the impact of environmental factors upon the quantum delocalization using the concurrence. \cite{Hill:1997iy, Wootters:1998dq} They visually demonstrated that smaller electronic coupling, larger reorganization energy, and higher temperature cause the dynamic localization,\cite{Renger:2004cz, Parson:2007fk} even in the case in which two electronic states resonate in a coupled homo-dimer. Recently, Ishizaki \cite{Ishizaki:2013jg} explored the influence of environmentally induced fluctuation timescales on the quantum mixing between electron donor and acceptor states in a model photo-induced electron transfer reaction. It was demonstrated that fast fluctuation and correspondingly fast solvation eradicates the quantum mixing between the donor and acceptor in the vicinity of the crossing point of the diabatic free energy surfaces, leading to a Marcus-type nonadiabatic reaction, whereas slow fluctuation sustains the quantum mixing and prompts the electron transfer reaction in an adiabatic fashion. \cite{Ishizaki:2013jg} Therefore, it is natural to question whether the nature of the quantum mechanically mixed electronic and vibrational states is altered through dynamic interactions with the environment.

The main purpose of this paper is to explore the impact of environmentally induced fluctuations upon the quantum mechanically mixed electronic and vibrational states in a simple electronic energy transfer system through electronic energy transfer dynamics and 2D electronic spectra calculations. The electronic energy transfer system used here is a coupled hetero-dimer, modeled on BChl 3 and 4 in the FMO protein. Further, we investigate the extent to which vibrational modes play a role in electronic energy transfer dynamics under the influence of the environmentally induced fluctuations.

\section{Model\label{sec:model}}
\setcounter{equation}{0}

To clarify the impact of the protein-induced dynamic fluctuations upon quantum mechanically mixed electronic-electronic and electronic-vibrational states in a systematic fashion, we consider the simplest electronic energy transfer system, a coupled hetero-dimer. To describe electronic energy transfer (EET), we restrict the electronic spectra of the $m$-th pigment in a pigment-protein complex (PPC) to the ground state, $\lvert \varphi_{mg} \rangle$, and the first excited state, $\lvert \varphi_{me} \rangle$, although higher excited states are sometimes of consequence in nonlinear spectroscopic signals.\cite{Brixner:2004ko,Bruggemann:2007gu} Thus, the model Hamiltonian of a pigment-protein complex comprising two pigments is expressed as
\begin{align}
	\hat{H}_{\rm PPC}
	=
	\sum_{m=1}^2
	\sum_{a=g,e}  \hat{H}_{ma}(x_m) 
	\lvert\varphi_{ma} \rangle\langle \varphi_{ma} \rvert
	+
	\sum_{m=1}^2
	\sum_{n=1}^2
	\hbar J_{mn} 
	\lvert  \varphi_{me} \rangle \langle \varphi_{mg} \rvert 
	\otimes
	\lvert \varphi_{ng} \rangle \langle \varphi_{ne} \rvert.
	\label{PPC-hamiltonian}
\end{align}
Here, $\hat{H}_{ma}(x_m)$ represents the diabatic Hamiltonian for the environmental and nuclear degrees of freedom (DOFs), $x_m$, when the system is in the $\lvert \varphi_{ma} \rangle$ state for $a=g,e$. The electronic coupling between the pigments, $\hbar J_{mn}$, may also be modulated by the environmental and nuclear DOFs. In the following, however, we assume that the nuclear dependence of $\hbar J_{mn}$ is vanishingly small and employ the Condon-like approximation as usual.

The Franck-Condon transition energy of the $m$-th pigment is obtained as
\begin{align}
	\hbar\Omega_m = \langle \hat{H}_{me}(x_m) - \hat{H}_{mg}(x_m) \rangle_{mg},
\end{align}
where the canonical average has been introduced, $\langle \dots \rangle_{ma} = \mathrm{Tr}(\dots \hat\rho_{ma}^{\rm eq})$ with the canonical distribution, $\hat\rho_{ma}^{\rm eq}=e^{-\beta\hat{H}_{ma}}/\mathrm{Tr}\,e^{-\beta\hat{H}_{ma}}$. Here, $\beta$ is the inverse temperature, $1/k_{\rm B}T$. The electronic energy of each diabatic state experiences fluctuations caused by the environmental and nuclear dynamics; these dynamics are described by the collective energy gap coordinate, such that
\begin{align}
	\hat{u}_m = \hat{H}_{me}(x_m) - \hat{H}_{mg}(x_m) - \hbar\Omega_m.
\end{align}
By definition, {\it the coordinate, $\hat{u}_m$, includes information associated not only with the electronic excited state, but also the electronic ground state.} In this work, we assume that the environmentally induced fluctuations can be described as Gaussian processes and that the relevant nuclear dynamics are harmonic vibrations. Under this assumption, the dynamic properties of the environmental and intramolecular vibrations are characterized by several types of two-body correlation functions of $\hat{u}_m(t) = e^{i\hat{H}_{mg}t/\hbar} \hat{u}_m e^{-i\hat{H}_{mg}t/\hbar}$, as shown below.

The environmental dynamics and intramolecular vibrational motion affecting the electronic transitions can be characterized by the nonequilibrium energy difference between the electronic ground and excited states as a function of a delay time, $t$, after the photoexcitation.\cite{Fleming:1996ky} The linear response theory allows one to express the nonequilibrium energy difference as 
\begin{align}
	\Delta E_m(t)
	=
	\hbar\Omega_m - \Psi_m(0) + \Psi_m(t).
\end{align}
Here, $\Psi_m(t)$ is the relaxation function \cite{Kubo:1985bs} defined as $\Psi_m(t) = \int^\infty_t ds\,\Phi_m(s)$ in terms of the response function of $\hat{u}_m(t)$, $\Phi_m(t) = \left\langle ({i}/{\hbar})[ \hat{u}_m(t), \hat{u}_m(0) ] \right\rangle_{mg}$. The relaxation function is independent of temperature, and the value of $\Delta E_m(0)-\Delta E_m(\infty)$ gives the Stokes shift magnitude.\cite{Fleming:1996ky} Under the assumption of Gaussian fluctuation and harmonic vibration, the Stokes shift is given as double the total reorganization energy and, hence, the relaxation function satisfies
\begin{align}
	\Psi_m(0) = 2 (\hbar\lambda_{m,{\rm env}} + \hbar\lambda_{m,{\rm vib}} ),
\end{align}
where $\hbar\lambda_{m,{\rm env}}$ and $\hbar\lambda_{m,{\rm vib}}$ denote the environmental and vibrational reorganization energies of the $m$-th pigment, respectively. The spectral density, $J_m(\omega)$, is defined as the imaginary component of the susceptibility\cite{Kubo:1985bs} and, hence, it is expressed in terms of $\Psi_m(t)$ as
\begin{align}
	J_m(\omega) 
	= 
	\omega\int^\infty_0 dt\, \Psi_m(t) \cos\omega t.
	\label{spectral-relaxation}
\end{align}
The fluctuations in the electronic transition energies are described by the symmetrized correlation function of $\hat{u}_m(t)$, defined as 
$D_m(t) = \left\langle ({1}/{2})\{ \hat{u}_m(t), \hat{u}_m(0) \}_+ \right\rangle_{mg}$.
In the quantum mechanical treatment, the Fourier transform of the symmetrized correlation function is expressed with that of the relaxation function, \cite{Kubo:1985bs} specifically, $D_m[\omega] = (\hbar\omega/2)\coth(\beta\hbar\omega/2) \Psi_m[\omega]$.
In the classical limit of $\coth(\beta\hbar\omega/2) \simeq 2k_{\rm B}T/\hbar\omega$, therefore, the correlation function of the fluctuations can be expressed in terms of $\Psi_m(t)$ as
\begin{align}
	\langle u_m(t) u_m(0) \rangle_{\rm cl} = k_{\rm B}T \cdot \Psi_m(t).
	\label{correlation-f-relaxation}
\end{align}
Equations~\eqref{spectral-relaxation} and \eqref{correlation-f-relaxation} suggest that the relaxation function, $\Psi_m(t)$, may be the key component for characterizing the relevant environmental dynamics and intramolecular vibrational motion.

The relaxation function can be separated into two components: an overdamped part originating from the environmental reorganization and an underdamped part induced by the intramolecular vibrations, i.e.,
\begin{align}
	\Psi_m(t) = \Psi_{m,{\rm env}}(t) + \Psi_{m,{\rm vib}}(t),
	\label{relaxation-f-decomp}
\end{align}
with $\Psi_{m,{\rm env}}(0) = 2\hbar\lambda_{m,{\rm env}}$ and $\Psi_{m,{\rm vib}}(0) = 2\hbar\lambda_{m,{\rm vib}}$. 
In order to focus on the timescale of the environmental dynamics affecting the electronic transition energies, we model the environmental component as an exponential decay form, with
\begin{align}
	\Psi_{m,{\rm env}}(t)
	=
	2\hbar\lambda_{\rm env} e^{-\gamma_{\rm env}t},
	\label{env-relax-func}
\end{align}
where $\gamma_{\rm env}^{-1}$ corresponds to the timescale of the environmental reorganization dynamics. Here, it should be noted that the initial behavior of Eq.~\eqref{env-relax-func} is coarse grained, as was commented in Ref.~\onlinecite{Ishizaki:2009jg}. Specifically, the value of $\partial_t\Psi_{m,{\rm env}}(t)\vert_{t=0}=-\Phi_{m,{\rm env}}(0)$ should vanish by definition. However, as long as finite timescales of the environmental dynamics are discussed, this coarse-grained nature of the exponential decay form does not cause fatal defects.
For simplicity, we consider a single intramolecular vibration on each of the pigments, with frequency, $\omega_{\rm vib}$, and the Huang-Rhys factor, $S$. We model the relaxation function originating from the vibration, $\Psi_{m,{\rm vib}}(t)$, using the Brownian oscillator model \cite{Mukamel:1995us} with the vibrational relaxation rate, $\gamma_{\rm vib}$, such that
\begin{align}
	\Psi_{m,{\rm vib}}(t)
	&=
	2\hbar\lambda_{\rm vib}
	\left(
		\cos \tilde\omega_{\rm vib} t
		+
    	\frac{\gamma_{\rm vib}}{\tilde\omega_{\rm vib}} \sin \tilde\omega_{\rm vib} t
	\right) 
	e^{-\gamma_{\rm vib} t},
	\label{vib-relax-func}
\end{align}
where $\tilde\omega_{\rm vib} = (\omega_{\rm vib}^2 - \gamma_{\rm vib}^2)^{1/2}$ and $\hbar\lambda_{\rm vib}=\hbar\omega_{\rm vib} S$ have been introduced.
Equation~\eqref{spectral-relaxation} gives the corresponding spectral density, which is reduced to
\begin{align}
	J_{m,{\rm vib}}(\omega) 
	= 
	\pi \hbar S \omega^2 [\delta(\omega-\omega_{\rm vib})-\delta(\omega+\omega_{\rm vib})],
	\label{spectral-density-delta}
\end{align}
in the slow vibrational relaxation limit, i.e., $\gamma_{\rm vib} \to 0$.

An adequate description of the EET dynamics is given by the reduced density operator, $\hat\rho_{\rm exc}(t)$, that is, the partial trace of the total PPC density operator, $\hat\rho_{\rm PPC}(t)$, over the environmental and nuclear DOFs: $\hat\rho_{\rm exc}(t) = \mathrm{Tr}_{\rm env}\hat\rho_{\rm PPC}(t)$. The Gaussian nature of the electronic excitation energy fluctuations, $\hat{u}_m$, allows a formally exact equation of motion to be obtained, that can describe the EET dynamics under the influence of the environmentally induced fluctuation and dissipation. The nature of these factors is described by Eq.~\eqref{relaxation-f-decomp}-\eqref{vib-relax-func}. Hence, the EET dynamics under the influence of the environmental/nuclear dynamics and the corresponding 2D electronic spectra can be described in a numerically accurate manner. The technical details of this calculation are given in Appendix~\ref{appendix-EOM} and Appendix~\ref{appendix-NR-response}.
In general, vibronic exciton states are introduced as energy eigenstates of the vibronic hamiltonian, which are independent of any environmental factors such as temperature and fluctuations. However, the correlation function approach in this work does not treat mixed electronic-vibrational states as the energy eigenstates, allowing us to describe the influence of environmentally induced fluctuations on the electronic-vibrational quantum mixture. The validity of employing the correlation functions to describe the quantum superposition between the electronic and vibronic transitions is demonstrated in Appendix~\ref{validity-of-theory}.

\section{Results and Discussion}
\setcounter{equation}{0}

In this section, we present and discuss the numerical results, in order to quantify the impact of the environment upon the quantum mechanically mixed electronic and vibrational states on the EET dynamics and  2D electronic spectra of a model dimer. We focus on a dimer which produces little beating of electronic origin in the absence of vibronic contributions.

The studied dimer is inspired by BChl 3 and 4 (pigments 1 and 2, respectively) in the FMO protein of {\it Chlorobaculum tepidum}, \cite{vanAmerongen:2000fy,Cho:2005bm,Adolphs:2006ey,Chenu:2013cf} which serve as the two lowest energy exciton states in the single-excitation manifold. The low-frequency vibrational modes of BChl in solutions or protein environments have been investigated in a number of experiments, demonstrating that the strongest vibrational mode is found at approximately $180\,{\rm cm^{-1}}$.\cite{Ratsep:2007fq,Ratsep:2011cq} Thus, to describe the effects of the Franck-Condon active vibrational modes, we consider a single vibrational mode with frequency $\omega_{\rm vib}=180\,{\rm cm^{-1}}$ in this work. For the sake of simplicity, the Franck-Condon transition energy of each pigment and their electronic coupling are set to $\Omega_1=12350\,{\rm cm^{-1}}$, $ \Omega_2 = \Omega_1 - 150\,{\rm cm^{-1}}$, and $J_{12} = -50\,{\rm cm^{-1}}$, respectively. The transition dipole moment directions are set to $\theta_{12} = 90^\circ$. In this situation, the gap between the two electronic energy eigenstates resonates with the vibrational frequency, $[{(\Omega_1-\Omega_2)^2+4J_{12}^2}]^{1/2} \simeq \omega_{\rm vib}$, and hence, the effects of the vibrational mode are expected to be maximized under the given conditions.\cite{Tiwari:2013dt} In general, the Huang-Rhys factors of BChl are small, \cite{Fransted:2012fs} and the factor associated with the vibrational mode in this work is set to $S=0.025$, \cite{Tiwari:2013dt} which is within the range of the experimentally measured values.\cite{Ratsep:2007fq,Ratsep:2011cq} The protein-environmental reorganization energy and reorganization time constant are set to $\lambda_{\rm env}=35{\rm \,cm^{-1}}$ and $\gamma_{\rm env}^{-1}=100\,{\rm fs}$,\cite{Brixner:2005wu,Adolphs:2006ey,Ishizaki:2009ky} and the vibrational relaxation rate is $\gamma_{\rm vib}^{-1} = 2\,{\rm ps}$.

\begin{figure}
    \includegraphics{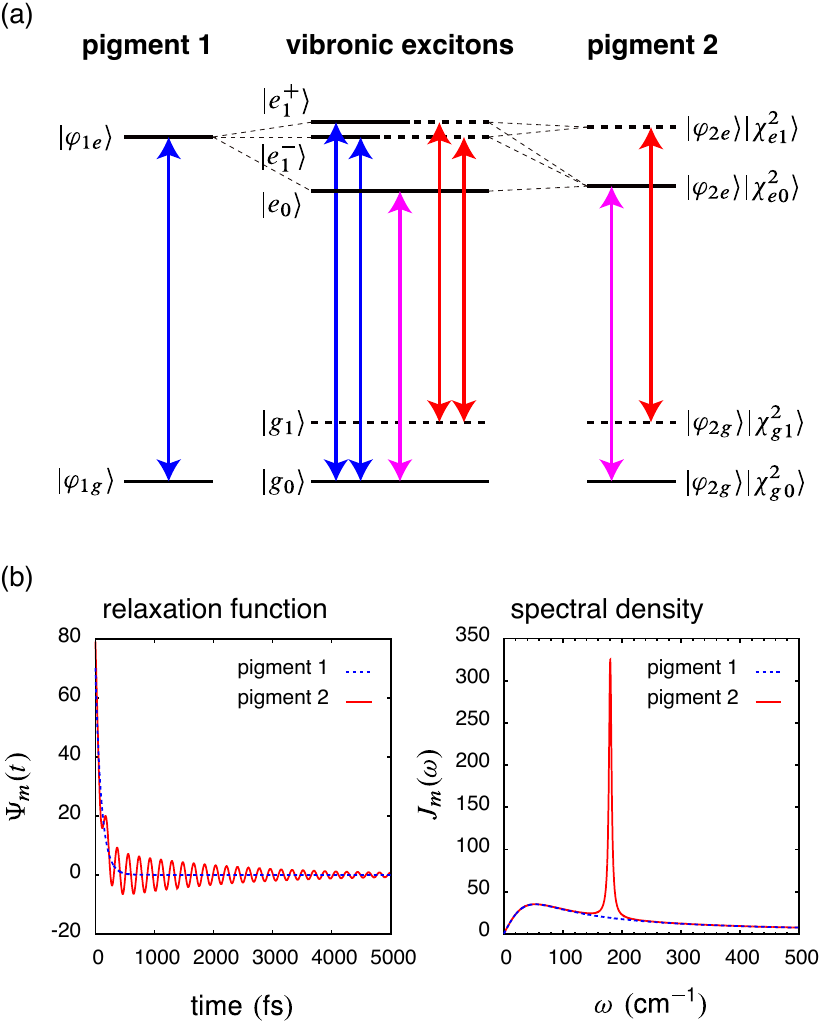}
	\caption{(a) Illustration of quantum mechanical mixing among vibronic transitions of pigments 1 and 2 in a weakly coupled hetero-dimer. Arrows are used to depict strong vibronic transitions only. For simplicity, only a single vibrational mode is considered in pigment 2. The exciton state, $\lvert e_0\rangle$, originates primarily from the vibronic state of pigment 2, $\lvert \varphi_{2e}\rangle\lvert \chi^2_{e0}\rangle$. Resonance or strong quantum mixing between the electronic state of pigment 1, $\lvert \varphi_{1e}\rangle$, and the vibronic state of pigment 2, $\lvert \varphi_{2e}\rangle\lvert \chi^2_{e1}\rangle$, creates the two vibronic excitons, $\lvert e_1^\pm \rangle$, in the absence of the environmentally induced fluctuation and dissipation. In the vibronic exciton picture, the overall ground state, $\lvert g_0 \rangle$, is given by $\lvert \varphi_{1g} \rangle\otimes \lvert \varphi_{2g}\rangle\lvert \chi^2_{g0} \rangle$, while $\lvert g_1 \rangle$ is identical to $\lvert \varphi_{2g} \rangle\lvert \chi^2_{g1} \rangle$.
	(b) The relaxation functions in Eqs.~\eqref{relaxation-f-decomp}-\eqref{vib-relax-func} and the corresponding spectral densities, Eq.~\eqref{spectral-relaxation}, employed for modeling the environmental dynamics and intramolecular vibrational motion affecting the electronic transitions of pigments 1 and 2.
	}
	\label{fig:vibronic-model}
\end{figure}

In what follows, we let $\lvert \chi_{av}^m \rangle$ denote the $v$-th vibrational level of the $a$-th electronic state in the $m$-th pigment. At low temperatures satisfying $\omega_{\rm vib} > k_{\rm B}T$, the system can be assumed in the $\lvert\varphi_{1g}\rangle \lvert \chi_{g0}^1\rangle \otimes \lvert\varphi_{2g}\rangle \lvert \chi_{g0}^2\rangle$ state before photoexcitation and, therefore, the two vibronic transitions, $\lvert\varphi_{1g}\rangle \lvert \chi_{g0}^1\rangle \leftrightarrow \lvert\varphi_{1e}\rangle \lvert \chi_{e0}^{1}\rangle$ and $\lvert\varphi_{2g}\rangle \lvert \chi_{g0}^2\rangle \leftrightarrow \lvert\varphi_{2e}\rangle \lvert \chi_{e1}^{2}\rangle$, can resonate and quantum mechanically mix. However, one should not overlook the fact that other resonances may be induced as time progresses after the photoexcitation. For example, once the $\lvert\varphi_{1e}\rangle \lvert \chi_{e0}^{1}\rangle$ state is created, it is possible for resonance to occur between $\lvert\varphi_{1g}\rangle \lvert \chi_{g1}^1\rangle \leftrightarrow \lvert\varphi_{1e}\rangle \lvert \chi_{e0}^{1}\rangle$ and $\lvert\varphi_{2g}\rangle \lvert \chi_{g0}^2\rangle \leftrightarrow \lvert\varphi_{2e}\rangle \lvert \chi_{e0}^{2}\rangle$. For the sake of simplicity, we consider the vibrational mode in pigment 2 only, as shown in Fig.~\ref{fig:vibronic-model}, and we focus on the resonance between the electronic transition, $\lvert \varphi_{1g}\rangle \leftrightarrow \lvert \varphi_{1e}\rangle$, of pigment 1 and the vibronic transition, $\lvert\varphi_{2g}\rangle \lvert \chi_{g0}^2\rangle \leftrightarrow \lvert\varphi_{2e}\rangle \lvert \chi_{e1}^{2}\rangle$, of pigment 2, which is induced by the coupling $J_{12} \langle \chi_{e1}^2 \vert \chi_{g0}^2 \rangle = -J_{12}\exp(-S/2)\sqrt{S}$.

\subsection{Population dynamics}

In this subsection, we explore the influence of intramolecular vibrations upon the spatial dynamics of electronic excitations in photosynthetic EET.

\begin{figure}
    \includegraphics{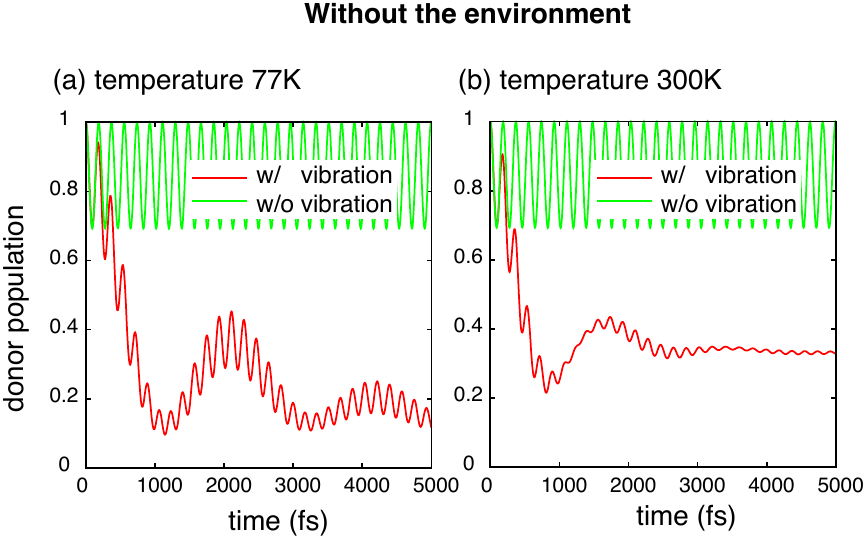}
	\caption{Time evolution of the energy donor population affected by intramolecular vibration of pigment 2 (acceptor) in the absence of protein environment effects at temperatures of 77 and 300\,K. In each panel, the red line represents the dynamics under the influence of the vibration, while the green line shows the dynamics without the vibration effects, i.e., the unitary time evolution. The vibrational frequency, the vibrational relaxation time constant, and the Huang-Rhys factor are set to $\omega_{\rm vib}=180\,{\rm cm^{-1}}$, $\gamma_{\rm vib}^{-1} = 2\,{\rm ps}$, and $S=0.025$, respectively.}
	\label{fig:pop-dynamics-noenv}
\end{figure}

In order to clarify the presence of the vibronic resonance and its influence upon the EET dynamics, Fig.~\ref{fig:pop-dynamics-noenv} presents the time evolution of the energy donor (pigment 1) population neglecting environmental factors at temperatures of (a) 77\,K and (b) 300\,K. In each panel, the red line represents the dynamics under the influence of the vibrations, whereas the green line gives the dynamics without the effects of the vibrations (i.e., the unitary time evolution) as a reference. Both panels indicate that intramolecular vibration dramatically affects the EET dynamics in the absence of the environmentally induced fluctuations and dissipation, despite the small Huang-Rhys factor of $S=0.025$. The population dynamics affected by the vibrations involve two oscillating components: faster oscillation with small amplitude and slower oscillation with large amplitude. By comparing the red and green lines, we can observe that the faster oscillating component indicates coherent EET between the electronic transition, $\lvert \varphi_{1g} \rangle \leftrightarrow \lvert \varphi_{1e} \rangle$, and the $0-0$ vibronic transition, $\lvert \varphi_{2g} \rangle\lvert \chi^2_{g0} \rangle \leftrightarrow \lvert \varphi_{2e} \rangle \lvert \chi^2_{e0} \rangle$, which is induced by the coupling constant, $J_{12} \langle \chi_{e0}^2 \vert \chi_{g0}^2 \rangle\simeq J_{12}\exp(-S/2) = 0.987 J_{12}$ for $S=0.025$. On the other hand, the slower oscillating component arises from the interaction between the electronic transition, $\lvert \varphi_{1g} \rangle \leftrightarrow \lvert \varphi_{1e} \rangle $, and the $0-1$ vibronic transition, $\lvert \varphi_{2g} \rangle\lvert \chi^2_{g0} \rangle \leftrightarrow \lvert \varphi_{2e} \rangle \lvert \chi^2_{e1} \rangle$, which is induced by the coupling constant, $J_{12} \langle \chi_{e1}^2 \vert \chi_{g0}^2 \rangle = - J_{12}\exp(-S/2)\sqrt{S} = - 0.156 J_{12}$. Moreover, the large amplitude of this slower component indicates a large mixing angle, which is the resonance between the electronic and vibronic transitions.

\begin{figure}
	\includegraphics{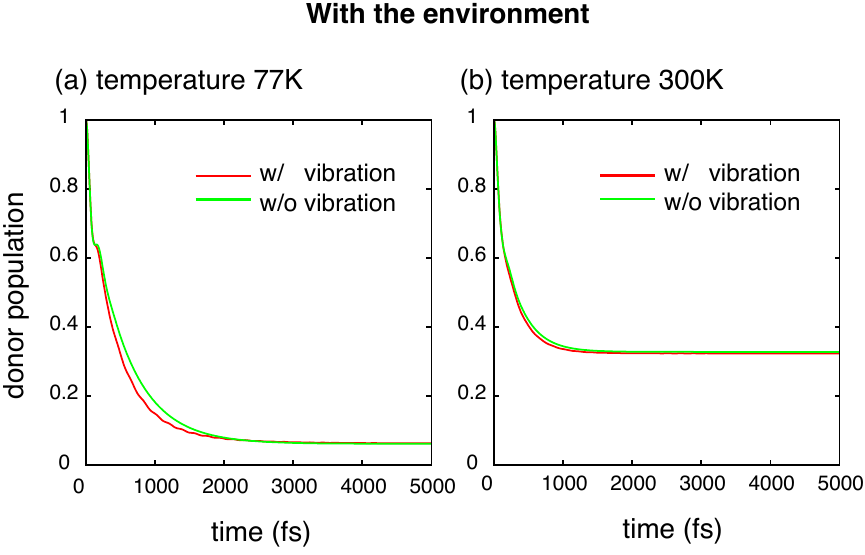}
	\caption{Time evolution of the energy donor population affected by intramolecular vibration of pigment 2 (acceptor) in the presence of environmental effects at temperatures of (a) 77 and (b) 300\,K. In each panel, the red line represents the dynamics under the influence of the vibration, whereas the green line shows the dynamics without the vibration effects. The environmental reorganization energy and the reorganization time constant are set to $\lambda_{\rm env}=35\,{\rm cm^{-1}}$ and $\gamma_{\rm env}^{-1} = 100\,{\rm fs}$, respectively. The other parameters are the same as in Fig.~\ref{fig:pop-dynamics-noenv}.}
	\label{fig:pop-dynamics-env}
\end{figure}

Figure~\ref{fig:pop-dynamics-env} shows the time evolutions of the energy donor (pigment 1) population affected by the intramolecular vibrations under the influence of the environmentally induced fluctuations and dissipation at temperatures of (a) 77\,K and (b) 300\,K. Similar to Fig.~\ref{fig:pop-dynamics-noenv}, the red lines represent the dynamics in the presence of the vibrations, whereas the green lines show the dynamics without the vibrations as a reference. The EET dynamics in Fig.~\ref{fig:pop-dynamics-env} show no wavelike motion in contrast with those in Fig.~\ref{fig:pop-dynamics-noenv}. However, it should be noted that the dynamics involve two distinct mechanisms. Comparing Fig.~\ref{fig:pop-dynamics-env} with Fig.~\ref{fig:pop-dynamics-noenv}, one recognizes that the short-term behaviors ($t\lesssim \gamma_{\rm env}^{-1}=100\,{\rm fs}$) arise from electronically coherent motion between the electronic transition of pigment 1 and the $0-0$ vibronic transition of pigment 2, and the subsequent hopping dynamics follow. As demonstrated in Ref.~\onlinecite{Ishizaki:2009jg}, the electronic excitation is delocalized just after the photoexcitation even in the incoherent hopping regime.  As time progresses, the dissipation of reorganization energy proceeds and the excitation becomes localized. The vibrational modes contribute to acceleration of the EET, albeit only slightly, because the vibronic resonance adds a new EET channel in comparison with the case without the vibrational mode. However, the overall behaviors of the EET dynamics are dominated by the environment, and the contribution of the vibrational modes decreases with increasing temperature. As can be seen in Eqs.~\eqref{correlation-f-relaxation} and \eqref{env-relax-func}, the root mean squared amplitude of the environmentally induced fluctuations in the electronic energy is given as $(2\hbar\lambda_{\rm env}k_{\rm B}T)^{1/2}$, which is evaluated as $61.2$ and $120.8\,{\rm cm^{-1}}$ at temperatures of 77 and 300\,K, respectively. On the other hand, the inter-site coupling constant inducing the vibronic resonance is evaluated as $J_{12} \langle \chi_{e1}^2 \vert \chi_{g0}^2 \rangle = 7.80\,{\rm cm^{-1}}$ for $S=0.025$. Hence, the inter-site coupling causing the vibronic resonance is one order of magnitude smaller than the fluctuation amplitudes and, therefore, the vibronic resonance does not play a significant role in the EET dynamics under the influence of the environmentally induced fluctuations.

\subsection{Beating in two-dimensional electronic spectra}

In this subsection, we discuss 2D electronic spectra, in order to explore the relationships between the spatial dynamics of the electronic excitations presented in the preceding subsection and the information content of the 2D electronic spectra.

\begin{figure}
	\includegraphics{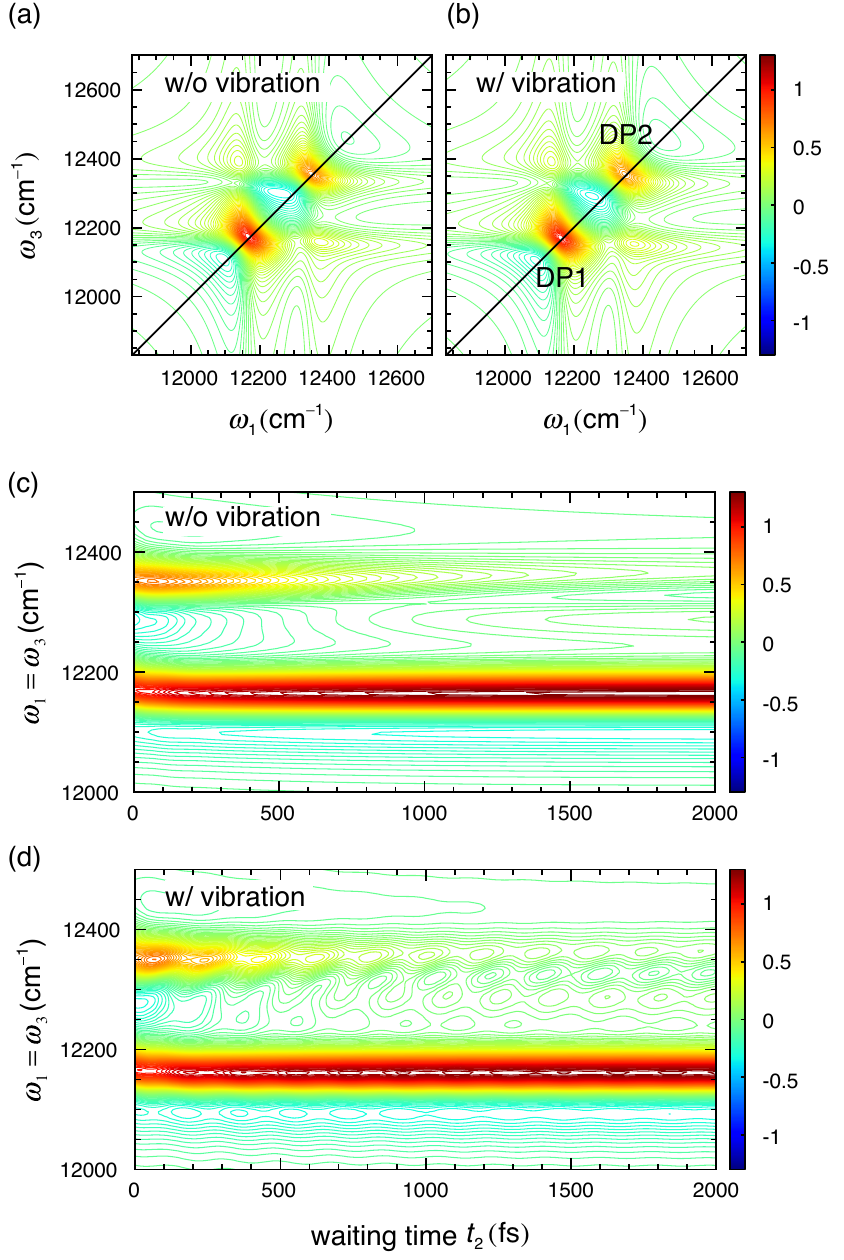}
	\caption{Nonrephasing 2D spectra at 77\,K of (a) a coupled-dimer without intramolecular vibration and (b) a coupled-dimer with intramolecular vibration. 
The waiting time is $t_2=0\, {\rm fs}$. The normalization of contour plots (a) and (b) is such that the maximum value of each spectrum is unity, and equally spaced contour levels ($0, \pm 0.02, \pm 0.04, \dots, $) are drawn. The lower-left and upper-right diagonal peaks are labeled DP1 and DP2, respectively. Panels (c) and (d) show the amplitudes of the diagonal cuts of 2D spectra (a) and (b) as functions of the waiting time, $t_2$, respectively. The calculations were performed with the parameters employed in Fig.~\ref{fig:pop-dynamics-env}.
}
	\label{fig:full2DNR-diagonal}
\end{figure}

Figures~\ref{fig:full2DNR-diagonal}a and \ref{fig:full2DNR-diagonal}b show calculated nonrephasing 2D spectra at 77\,K without and with the intramolecular vibration, respectively. These calculations were performed using the same equation of motion and parameters as in Fig.~\ref{fig:pop-dynamics-env}. The lower-left and upper-right diagonal peaks are labeled DP1 and DP2, respectively. The amplitudes of the diagonal cuts are also presented as a function of the waiting time, $t_2$, in Fig.~\ref{fig:full2DNR-diagonal}c and \ref{fig:full2DNR-diagonal}d. Although the EET dynamics in Fig.~\ref{fig:pop-dynamics-env} and the 2D spectra at $t_2=0$ do not exhibit a significant change originating from the vibrational mode, Figs.~\ref{fig:full2DNR-diagonal}c and \ref{fig:full2DNR-diagonal}d exhibit a qualitative difference. Figure~\ref{fig:full2DNR-diagonal}c shows monotonous decays of the diagonal peaks, whereas Fig.~\ref{fig:full2DNR-diagonal}d exhibits peak beating persisting for up to a minimum of 2\,ps. Apparently, this beating is caused by the presence of the vibrational mode.

\begin{figure}
	\includegraphics{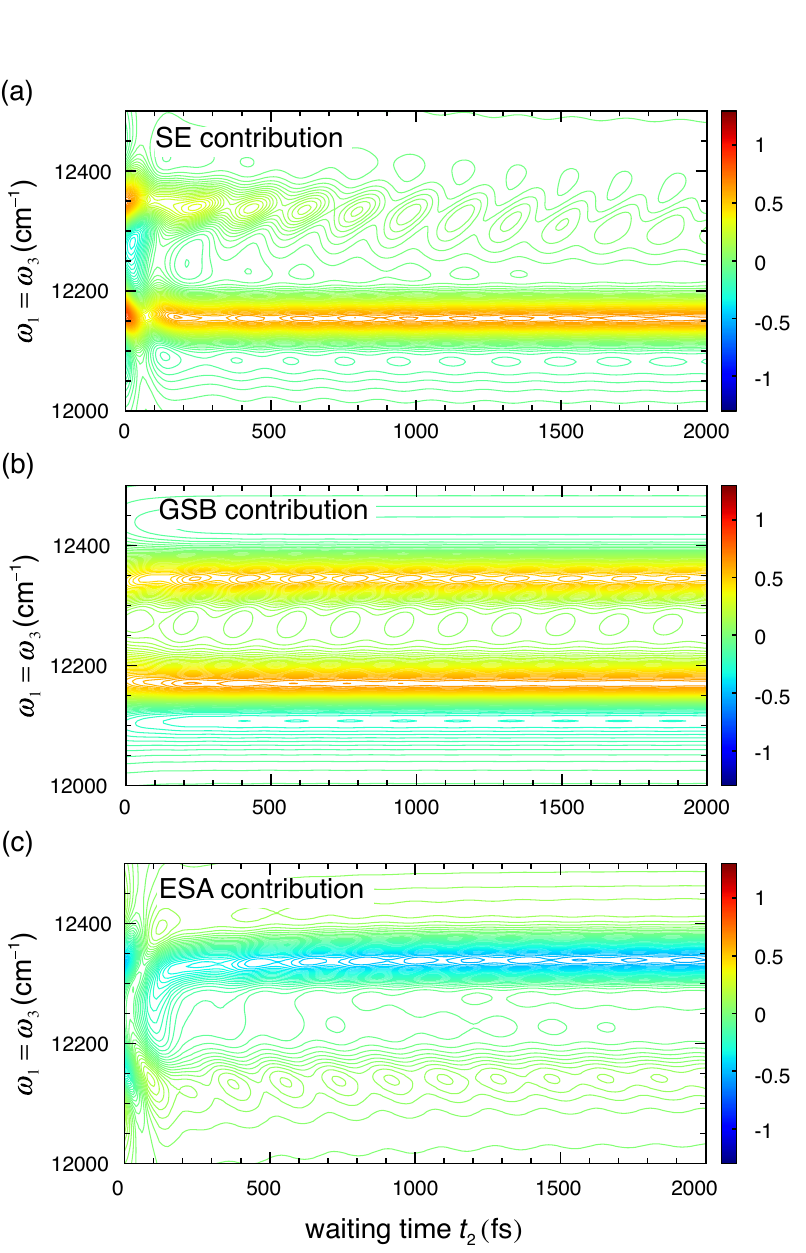}
	\caption{
	Time evolutions of the diagonal cuts of (a) the stimulated emission, (b) the ground state bleaching, and (c) the excited state absorption contributions to the nonrephasing 2D spectra given in Figs.~\ref{fig:full2DNR-diagonal}b and \ref{fig:full2DNR-diagonal}d. The normalization of the contour plots is such that the maximum value of Fig.~\ref{fig:full2DNR-diagonal}b is unity, and equally spaced contour levels ($0, \pm 0.02, \pm 0.04, \dots, $) are drawn.
	}
	\label{fig:diagonal-cut}
\end{figure}

\begin{figure}
	\includegraphics{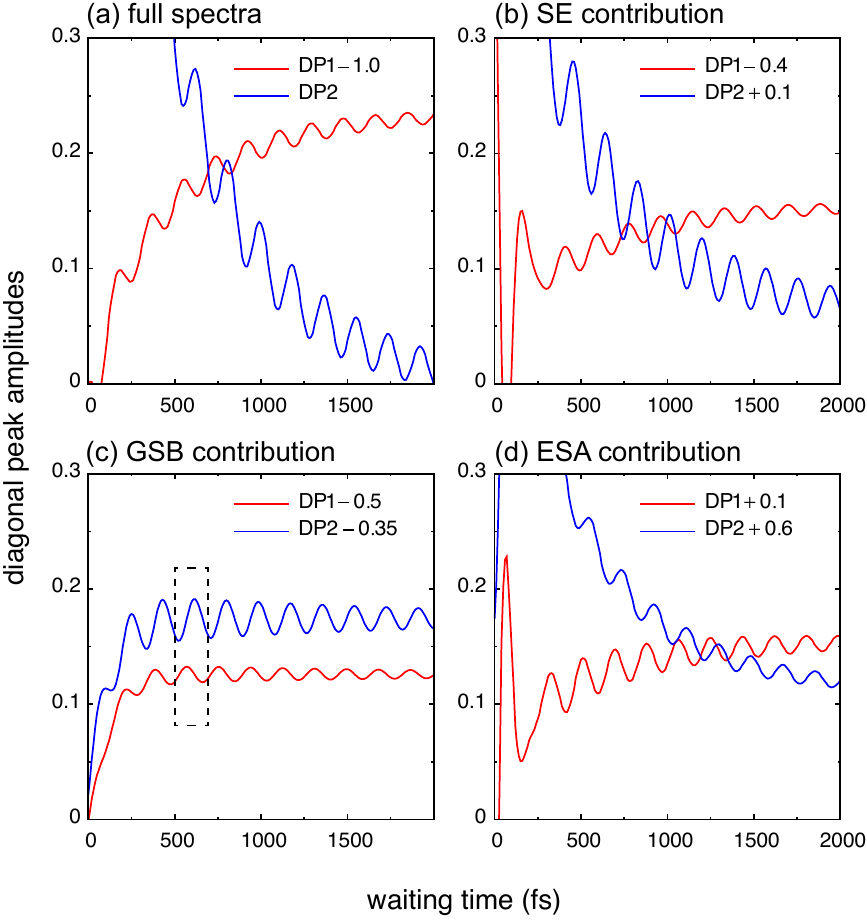}
	\caption{
	Time evolutions of the diagonal peak amplitudes, DP1 $(\omega_1=\omega_3=12,166\,{\rm cm^{-1}})$ and DP2 $(\omega_1=\omega_3=12,350\,{\rm cm^{-1}})$, of (a) the nonrephasing 2D spectrum, (b) the stimulated emission contribution, (c) the ground state bleaching contribution, and (d) the excited state absorption contribution. The normalization of the plots is such that the maximum value of Fig.~\ref{fig:full2DNR-diagonal}b is unity. In order to clarify the relationship between the beating amplitude magnitudes in the individual contributions, the peak amplitudes are shifted.
	}
	\label{fig:peaks-decomposition}
\end{figure}

To clarify the origins of the beatings in Fig.~\ref{fig:full2DNR-diagonal}d, we decompose the nonrephasing 2D spectra to the stimulated emission (SE), ground state bleaching (GSB), and excited state absorption (ESA) contributions. Figure~\ref{fig:diagonal-cut} shows the time-evolutions of the diagonal cuts of the individual contributions. The time-evolutions of the amplitudes of the diagonal peaks, DP1 $(\omega_1=\omega_3=12,166\,{\rm cm^{-1}})$ and DP2 $(\omega_1=\omega_3=12,350\,{\rm cm^{-1}}$) in the contributions are also shown in Fig.~\ref{fig:peaks-decomposition}. The SE and ESA contributions exhibit beating corresponding to quantum superposition between the vibronic exciton states, $\lvert e_0 \rangle$ and $\lvert e_1^{\pm} \rangle$, in Fig.~\ref{fig:vibronic-model}, whereas the beating in the GSB contribution arises from quantum superposition between the vibrational states in pigment 2, $\lvert g_0 \rangle$ and $\lvert g_1 \rangle$. As shown in Figs.~\ref{fig:diagonal-cut} and \ref{fig:peaks-decomposition}, the SE, GSB, and ESA pathways exhibit similar contribution levels to the diagonal peak beatings in the nonrephasing 2D spectra, although the SE and ESA pathways exhibit larger beating amplitudes in the short-term region ($\sim 300\,{\rm fs}$), which originate from the electronic coherence in the single-excitation manifold.
The present observation corroborates the recent theoretical argument by Plenio, Almeida, and Huelga,\cite{Plenio:2013bg} which suggested that the SE, GSB, and ESA contributions in the rephasing 2D spectra could be of the same order in realistic parameter ranges with the use of a simplified Markovian quantum master equation.

\begin{figure}
	\includegraphics{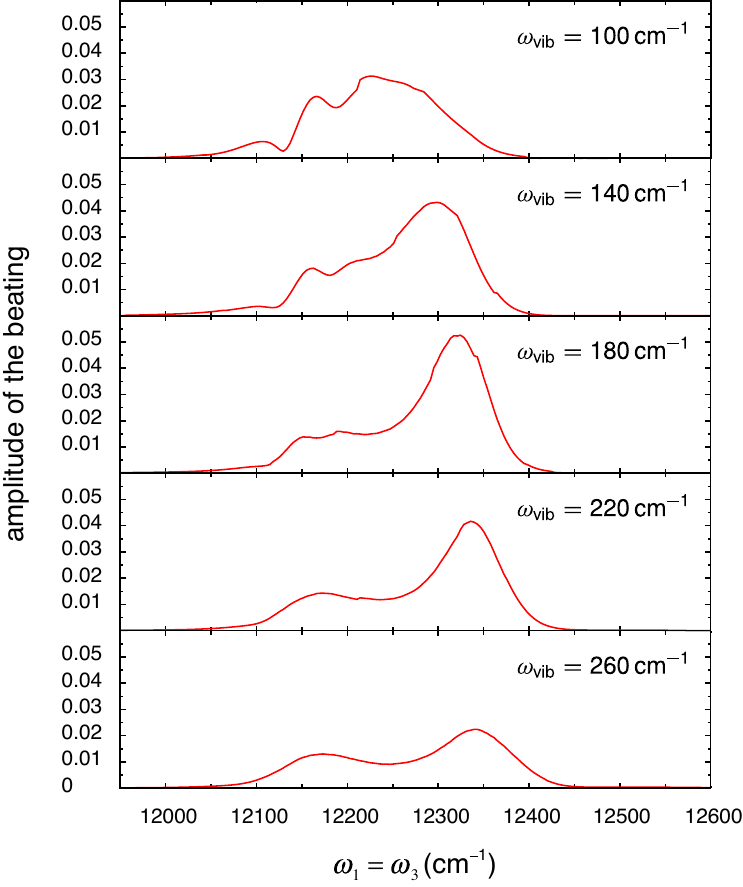}
	\caption{Beating amplitude distribution along the diagonal line of the GSB contribution in the nonrephasing 2D spectra for various values of the vibrational frequency, $\omega_{\rm vib}$. 
	The temperature is $77\,{\rm K}$.
	The normalization of the individual plots is such that the maximum value of Fig.~\ref{fig:full2DNR-diagonal}b is unity.
	}
	\label{fig:resonance-vib.1}
\end{figure}

In order to clarify the extent to which the quantum mixture between the electronic transition of pigment 1 and the vibronic transitions of pigment 2 contributes to the observed beating in the nonrephasing 2D spectra, we explore the effects of the environmentally induced fluctuations on beating amplitude. 
For this purpose, we address beating in the GSB pathways, that is, the vibrational coherence between $\lvert g_0 \rangle$ and $\lvert g_1 \rangle$ in the electronic ground state. This is because the GSB pathway is unaffected by the energy transfer processes in the single-excitation manifold and, thus, it allows one to investigate interplays between the quantum mixture and the environmentally induced fluctuations. The beating amplitudes in the GSB contribution observed in the nonrephasing 2D spectra are determined as being the difference between the first and second extremal values after $t_2=500\,{\rm fs}$, as indicated by the dashed-line box in Fig.~\ref{fig:peaks-decomposition}c, where the environmental reorganization dynamics have already been completed and the beating characteristics can be extracted cleanly. Figure~\ref{fig:resonance-vib.1} shows the beating amplitude along the diagonal line of the GSB contribution for various values of $\omega_{\rm vib}$. From the plots, we find two characteristic beating behaviors in the GSB contribution of the nonrephasing 2D spectra. 
Firstly, a local maximum of the beating amplitude is observed at the DP1 position ($\omega_1 = \omega_3=12,166\,{\rm cm^{-1}}$), independently of the vibrational frequency. This beating is caused by the ground state vibrational coherence created via the lower exciton, $\lvert e_0 \rangle$, that is, through the Liouville pathway of $\lvert g_0 \rangle\langle g_0 \rvert \to \lvert e_0 \rangle\langle g_0 \rvert \to \lvert g_1 \rangle\langle g_0 \rvert \to \lvert e_0 \rangle\langle g_0 \rvert \to \lvert g_0 \rangle\langle g_0 \rvert$.
Secondly, another local maximum is found at various locations in the vicinity of DP2 ($\omega_1=\omega_3=12,350\,{\rm cm^{-1}}$), depending on the vibrational frequency. This beating originates from the ground state vibrational coherence created via the upper vibronic excitons, $\lvert e_1^{\pm} \rangle$, that is, through the Liouville pathways of $\lvert g_0 \rangle\langle g_0 \rvert \to \lvert e_1^{\pm} \rangle\langle g_0 \rvert \to \lvert g_1 \rangle\langle g_0 \rvert \to \lvert e_1^{\pm} \rangle\langle g_0 \rvert \to \lvert g_0 \rangle\langle g_0 \rvert$. The transition energy of the vibronic excitons, $\lvert e_1^{\pm} \rangle$, differs according to the vibrational frequency and, hence, the location of the beating amplitude maxima vary. Moreover, the mixing angle between the electronic transition of pigment 1 and the vibronic transition of pigment 2 depends on the vibration frequency, which causes variations in the beating amplitudes. This fact leads to the concept that the degree of quantum mixing between the electronic transition of pigment 1 and the vibronic transition of pigment 2 may be visualized as the beating amplitude of the ground state vibrational coherence.

\begin{figure}
	\includegraphics{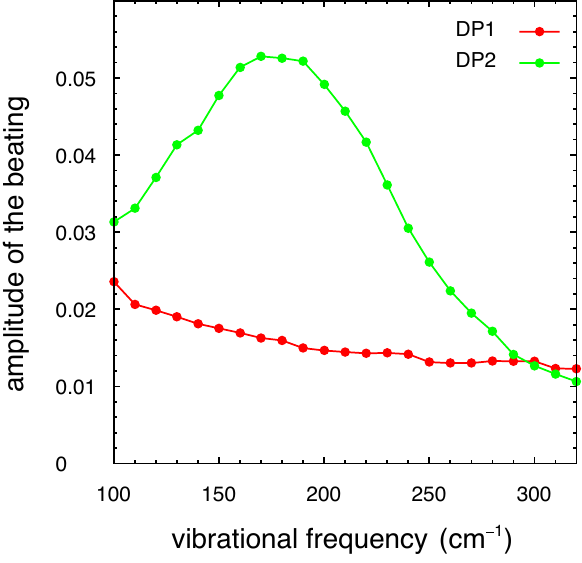}
	\caption{Beating amplitude of DP1 $(\omega_1 = \omega_3=12,166\,{\rm cm^{-1}})$ and the local maximum value of the beating amplitude in the vicinity of DP2 $(\omega_1=\omega_3=12,350\,{\rm cm^{-1}})$ as a function of the vibrational frequency, $\omega_{\rm vib}$.
	The temperature is $77\,{\rm K}$.
	The normalization of the individual plots is such that the maximum value of Fig.~\ref{fig:full2DNR-diagonal}b is unity.
	}
	\label{fig:resonance-vib.2}
\end{figure}

Figure~\ref{fig:resonance-vib.2} plots the beating amplitude of DP1 and the local maxima values of the beating amplitude in the vicinity of DP2 as a function of $\omega_{\rm vib}$. The beating at the DP1 location originates primarily from the ground state vibrational coherence created via $\lvert e_0 \rangle$, which is minimally dependent on the quantum mixture between the electronic and vibronic transitions. However, the beating amplitude of DP1 decreases with the increasing vibrational frequency. This behavior is explained as being due to the interference between the beating of DP1 and that in the vicinity of DP2, as shown in Fig.~\ref{fig:resonance-vib.1}. On the other hand, the beating amplitude near DP2 exhibits a maximum in the vicinity of $\omega_{\rm vib}=180\,{\rm cm^{-1}}$. These findings reflect the degree of quantum mixing and, thus, the degree of intensity borrowing. Tiwari, Peters, and Jonas \cite{Tiwari:2013dt} demonstrated that the beating amplitude of a cross-peak in the rephrasing GSB pathway may be enhanced over a wide vibrational frequency range around the vibronic resonance. Their finding is consistent with those of the present study that adequately describes the impact of the environmentally induced fluctuation and dissipation upon the quantum mixing. Figure~\ref{fig:resonance-vib.2} implies that the quantum mixing between the vibronic transitions in BChl in the FMO protein, which induces vibronic coherence in the electronic ground and excited states, can be rather robust. This holds even under the influence of the environmentally induced fluctuations, at least at cryogenic temperatures.

\begin{figure}
	\includegraphics{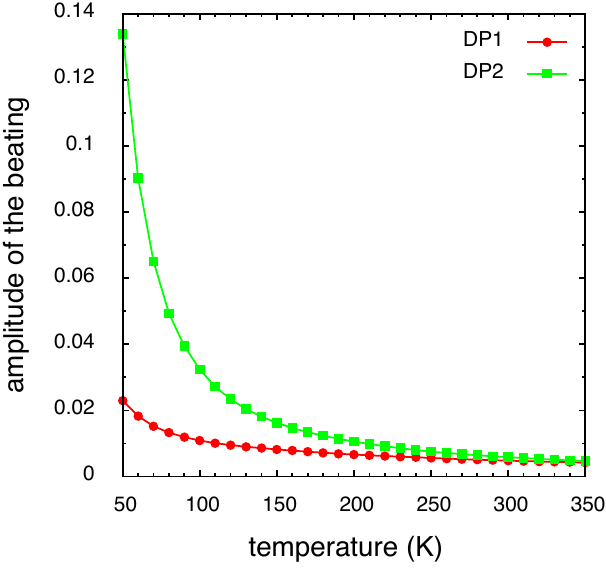}
	\caption{Beating amplitude of DP1 $(\omega_1 = \omega_3=12,166\,{\rm cm^{-1}})$ and the local maximum value of the beating amplitude in the vicinity of DP2 $(\omega_1=\omega_3=12,350\,{\rm cm^{-1}})$ as a function of temperature. The vibrational frequency is fixed to $\omega_{\rm vib}=180\,{\rm cm^{-1}}$. The normalization of the individual plots is such that the maximum value of Fig.~\ref{fig:full2DNR-diagonal}b is unity.
	}
	\label{fig:T_dependence}
\end{figure}

\begin{figure}
	\includegraphics{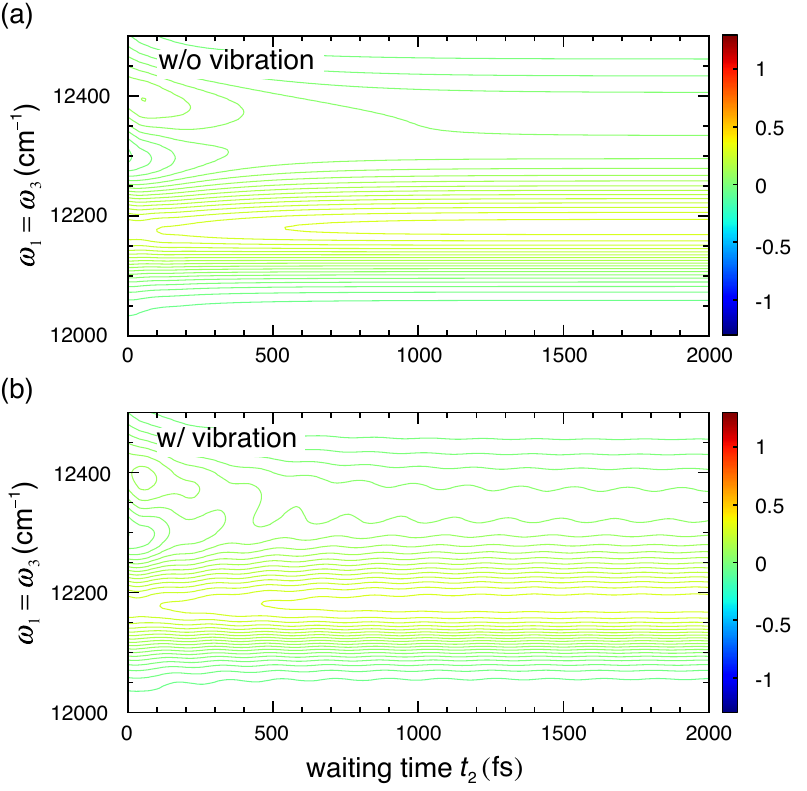}
	\caption{Time evolutions of the diagonal cut of the nonrephasing 2D spectra at 300\,K of (a) a coupled-dimer without intramolecular vibration and (b) a coupled-dimer with intramolecular vibration. The normalization of contour plots (a) and (b) is such that the maximum values of Fig.~\ref{fig:full2DNR-diagonal}a and \ref{fig:full2DNR-diagonal}b are unity. Equally spaced contour levels ($0, \pm 0.02, \pm 0.04, \dots, $) are drawn.
	}
	\label{fig:diagonal-temperature-300K}
\end{figure}

However, the root mean squared amplitude of the environmentally induced fluctuations in the electronic energy is estimated as $(2\hbar\lambda_{\rm env}k_{\rm B}T)^{1/2}$ using Eqs.~\eqref{correlation-f-relaxation} and \eqref{env-relax-func}, and the quantum mixing between the strong electronic and weak vibronic transitions may be more fragile against the protein-induced fluctuations at higher temperatures. Figure~\ref{fig:T_dependence} shows the beating amplitude of DP1 and the local maximum value of the beating amplitude in the vicinity of DP2 as a function of temperature for $\omega_{\rm vib}=180\,{\rm cm^{-1}}$. The beating of DP1 originates from the ground state vibrational coherence created via $\lvert e_0 \rangle$, independently of the quantum mixture and, thus, it exhibits minimal temperature dependence as well as the vibrational frequency dependence apparent in Fig.~\ref{fig:resonance-vib.2}. However, the beating amplitude in the DP2 region is strongly dependent on temperature. This indicates that the environmentally induced fluctuations of larger amplitude at higher temperature eradicate the quantum mixing between the strong electronic transition of pigment 1 and the weak vibronic transition of pigment 2. At the high temperature limit, the quantum mixture and, thus, the intensity borrowing mechanism, diminish. Hence, the beating amplitudes of DP1 and DP2 converge towards each other. Figure~\ref{fig:diagonal-temperature-300K} shows the time-evolutions of the amplitudes of the diagonal cuts of nonrephasing 2D electronic spectra at 300\,K for (a) a coupled dimer without the intramolecular vibration in pigment 2 and (b) a coupled dimer with the intramolecular vibration. The intensity borrowing mechanism is eradicated at 300\,K, as indicated in Fig.~\ref{fig:T_dependence}, and the vibrational contribution in Fig.~\ref{fig:diagonal-temperature-300K}b almost vanishes as a result.

The fluorescence line-narrowing measurement of the FMO protein reveals that a vibrational mode at $117\,{\rm cm^{-1}}$ and several modes between $167-202\,{\rm cm^{-1}}$ have relatively strong Huang-Rhys factor magnitudes.\cite{Ratsep:2007fq} As shown in Fig.~\ref{fig:resonance-vib.2}, the beating amplitude in the GSB contribution is enhanced over a wide range of vibrational frequencies near $\omega_{\rm vib}=180\,{\rm cm^{-1}}$ at cryogenic temperature, i.e., 77\,K. This behavior supports the recent theoretical result reported by Tempelaar, Jansen, and Knoester,\cite{Tempelaar:2014vu} who showed that the vibrational modes of $117\,{\rm cm^{-1}}$ and $185\,{\rm cm^{-1}}$ induce long-lasting beating of the exciton $1-3$ cross peak in the rephasing 2D spectra of the FMO protein. This was also investigated by Panitchayangkoon {\it et al.}\cite{Panitchayangkoon:2010fw} However, the amplitudes of the beats induced by the electronic-vibrational quantum mixture decrease with increasing temperature, as shown in Figs.~\ref{fig:T_dependence}. Moreover, the vibrational contribution almost vanishes, as shown in Fig.~\ref{fig:diagonal-temperature-300K}. Panitchayangkoon {\it et al.}\cite{Panitchayangkoon:2010fw} showed that quantum beats in the FMO complex persist for a minimum of 1.5\,ps and 300\,fs at 77 and 277\,K, respectively, whereas Ishizaki and Fleming \cite{Ishizaki:2009ky} theoretically predicted that electronic coherence in the FMO complex could persist for 700 and 300\,fs at 77 and 300\,K, respectively. The experimental data showed good agreement with the theoretical prediction regarding the electronic coherence at physiological temperature; however, the experimental data exhibited a significantly longer quantum beat lifetime in comparison with the theoretical result. Hence, the temperature dependence presented in Figs.~\ref{fig:full2DNR-diagonal}, \ref{fig:T_dependence}, and \ref{fig:diagonal-temperature-300K} indicates that the long-term  behavior of the experimentally observed long-lasting quantum beats is primarily vibrational in nature.

\section{Concluding Remarks}
\setcounter{equation}{0}

In this work, we have explored the impact of environmentally induced fluctuations upon the quantum mechanically mixed electronic and vibrational states of pigments in terms of EET dynamics and the corresponding 2D electronic spectra. For this purpose, we have addressed a model dimer extracted from the FMO protein and have performed numerical calculations of EET dynamics and the corresponding 2D electronic spectra in an accurate manner.

At cryogenic temperature, the resonance between the strong $0-0$ and weak $0-1$ vibronic pigment transitions are rather robust, even under the influence of protein-induced fluctuations, despite the use of a small Huang-Rhys factor, i.e., $S=0.025$. This resonance results in quantum beats of vibrational origin in both the electronic ground and excited states in the 2D electronic spectra via the intensity borrowing mechanism. This is not specific to the fine resonance between the vibronic transitions, and the beating originating from the vibrational coherence can be enhanced over a rather wide range of vibrational frequencies around the resonance condition, at cryogenic temperature at least. However, the amplitudes of the environmentally induced fluctuations in the electronic excitation energy increase with increasing temperature. The present calculations demonstrate that the quantum mixing between the strong $0-0$ and $0-1$ vibronic transitions is more fragile against fluctuations at higher temperatures, and the quantum beats originating from the vibrational coherence in 2D electronic spectra almost vanish at physiological temperatures.

Furthermore, we have investigated EET dynamics using the same dynamic equation and parameters as in the calculations for the 2D electronic spectra. In the absence of the environmental effects, the resonance between the vibronic transitions dramatically alters the EET dynamics, despite the small $S~(=0.025)$. However, under the influence of the environmentally induced fluctuations and dissipation, the vibronic resonance contributes only slightly to acceleration of the electronic energy transfer, both at cryogenic and physiological temperatures, although the 2D electronic spectra clearly exhibit quantum beating that is enhanced by the vibronic resonance at cryogenic temperature. One of the remarkable benefits of nonlinear multidimensional spectroscopic techniques is their sensitivity to details of the systems and dynamics under investigation. The calculated 2D electronic spectra in the present study indicate vanishingly small effects of the vibronic quantum mixing upon the EET dynamics and present them as long-lived quantum beating. In other words, this suggests that the concept that oscillatory behavior in 2D data provides genuinely relevant information on the systems and dynamics under investigation should be carefully considered. Recently, Fuller {\it et al.} \cite{Fuller:2014iz} and Romero {\it et al.} \cite{Romero:2014jm} revealed the presence of long-lived quantum beats in the photosystem II reaction center through the use of 2D electronic spectroscopy. Both works demonstrate that the observed long-lived beats are induced by the electronic and vibrational resonance between the electronic exciton and the primary charge transfer states, and they suggest that the electronic-vibrational resonance may represent an important design principle for enabling the high quantum efficiency of charge separation in the reaction center. These findings are indeed intriguing. However, the reorganization energies and, thus, the protein-induced fluctuations associated with charge transfer states are generally large in comparison to those for the electronic exciton states and, hence, further investigation of the influence of the electronic-vibrational resonance upon the charge transfer dynamics is required. This will be the subject of future study.

\begin{acknowledgments}
This work was supported by Grants-in-Aid for Scientific Research (Grant Number 25708003) from the Japan Society for the Promotion of Science and the U.S. Department of Energy, Office of Science, Office of Basic Energy Sciences, Chemical Sciences, Geosciences, and Biosciences Division.
\end{acknowledgments}

\appendix

\section{Equation of motion\label{appendix-EOM}}
\renewcommand{\theequation}{\ref{appendix-EOM}.\arabic{equation}}
\setcounter{equation}{0}

In order to derive the equation of motion to describe EET dynamics, we organize the product states in order of elementary excitation number. The overall ground state with zero excitation reads $\lvert g \rangle = \lvert \varphi_{1g} \rangle \lvert \varphi_{2g} \rangle$. The presence of a single excitation at pigment 1 is expressed as $\lvert 1 \rangle = \lvert \varphi_{1e}\rangle\lvert \varphi_{2g}\rangle$, whereas the other is $\lvert 2 \rangle = \lvert \varphi_{1g}\rangle\lvert \varphi_{2e}\rangle$. Since the intensity of sunlight is weak, the single-excitation manifold is of primary importance under physiological conditions. However, nonlinear spectroscopic techniques such as 2D electronic spectroscopy can populate some higher excitation manifolds, e.g., the double-excitation manifold comprising $\lvert 12 \rangle = \lvert \varphi_{1e}\rangle\lvert \varphi_{2e}\rangle$. Hence, the PPC Hamiltonian in Eq.~\eqref{PPC-hamiltonian} can be recast as
\begin{align}
	\hat{H}_{\rm PPC}
	=
	\hat{H}_{\rm exc} + \hat{H}_{\rm env} + \hat{H}_{\rm exc-env},
	\label{decomposed-hamiltonian}
\end{align}
with
\begin{align}
	\hat{H}_{\rm exc}
	&=
	E_g \lvert g \rangle\langle g \rvert 
	\notag\\
	&+ 
	\sum_{m=1}^2 (E_g + \hbar\Omega_m) 
	\lvert m \rangle \langle m \rvert 
	+
	\sum_{m=1}^2 \sum_{n \ne m}
	\hbar J_{mn} 
	\lvert m \rangle\langle n \rvert 
	\notag\\
	&+
	(E_g + \hbar\Omega_1 + \hbar\Omega_2) 
	\lvert 12 \rangle \langle 12 \rvert,
\\
	\hat{H}_{\rm env}
	&=
	\sum_{m=1}^2 \hat{H}_{mg}(x_m) - E_g, 
\\
	\hat{H}_{\rm exc-env}
	&=
	\sum_{m=1}^2 \hat{u}_m \lvert m \rangle \langle m \rvert 
	+ 
	(\hat{u}_1 + \hat{u}_2) \lvert 12 \rangle \langle 12 \rvert.
\end{align}
In the above, the excitation-vacuum energy, $E_g = \sum_m \langle \hat{H}_{mg}(x_m) \rangle_{mg}$, has been introduced; however, we set $E_g = 0$ without loss of generality. In order to derive the reduced density operator to describe the EET dynamics, we suppose that the total PPC system at the initial time, $t=0$, is in the factorized product state of the form, $\hat\rho_{\rm PPC}^{\rm eq}=\hat\rho_{\rm exc} \prod_{m} \hat\rho_{mg}^{\rm eq}$. This factorized initial condition is generally considered to be unphysical in the literature on open quantum systems,\cite{Grabert:1988ha} since it neglects the inherent correlation between a system of interest and its environment. In electronic excitation processes, however, this initial condition is of no consequence, because it corresponds to the electronic ground state or an electronic excited state generated in accordance with the vertical Franck-Condon transition. The reduced density operator for the electronic excitation, $\hat\rho_{\rm exc}(t)$, is given by $\hat\rho_{\rm exc}(t) = \mathrm{Tr}_{\rm env}[\hat{G}(t) \hat\rho_{\rm PPC}^{\rm eq} ]$, where $\hat{G}(t)$ is the retarded propagator of the total PPC system in the Liouville space. The Gaussian property of $\hat{u}_m$ enables one to derive a formally exact equation of motion as\cite{Ishizaki:2010fx}
\begin{align}
	\frac{d}{dt}\tilde\rho_{\rm exc}(t)
	&=
	\sum_{m=1}^2 
	\mathrm{T}_+
	\int^t_0 ds \tilde{K}_m^{(1)}(t,s)\tilde\rho_{\rm exc}(t)
	+
	\mathrm{T}_+
	\int^t_0 ds \tilde{K}_{12}^{(2)}(t,s)\tilde\rho_{\rm exc}(t),
	\label{EOM}	
\end{align}
where the interaction representation has been employed with respect to $\hat{H}_{\rm exc}+\hat{H}_{\rm env}$. The integration kernel for the single-excitation manifold, $\tilde{K}_m^{(1)}(t,s)$, is given by
\begin{align}
	\tilde{K}_m^{(1)}(t,s)
	=
	-\frac{1}{\hbar^2} \tilde{V}_m(t)^\times
	\left[
		D_m(t-s) \tilde{V}_m(s)^\times - i\frac{\hbar}{2}\Phi_m(t-s) \tilde{V}_m(s)^\circ
	\right],
	\label{kernel1}
\end{align}
where $\hat{V}_m = \lvert m \rangle\langle m \rvert$ has been introduced, and $D_m(t)$ and $\Phi_m(t)$ are the symmetrized correlation function and the response function of the collective energy gap coordinate $\hat{u}(t)$, respectively.
On the other hand, the integration kernel used to describe the double-excitation manifold, $\tilde{K}_m^{(2)}(t,s)$, is given by
\begin{align}
	\tilde{K}_{12}^{(2)}(t,s)
	=
	-\frac{1}{\hbar^2} \tilde{V}_{12}(t)^\times
	\left[
		D_{12}(t-s) \tilde{V}_{12}(s)^\times - i\frac{\hbar}{2}\Phi_{12}(t-s) \tilde{V}_{12}(s)^\circ
	\right],
	\label{kernel2}
\end{align}
with $D_{12}(t) = D_1(t)+D_2(t)$ and $\Phi_{12}(t) = \Phi_1(t)+\Phi_2(t)$. The superoperator notation has been introduced, $\hat{O}_1^\times \hat{O}_2 = \hat{O}_1\hat{O}_2-\hat{O}_2\hat{O}_1$ and $\hat{O}_1^\circ \hat{O}_2 = \hat{O}_1\hat{O}_2+\hat{O}_2\hat{O}_1$, for any operators $\hat{O}_1$ and $\hat{O}_2$. In Eq.~\eqref{EOM}, the chronological time ordering operator, ${\rm T}_+$, resequences and mixes the superoperators, $\tilde{V}_m(t)^\times$ and $\tilde{V}_m(t)^\circ$, comprised in $\tilde{K}_m^{(n)}(t,s)$ and $\tilde\rho_{\rm exc}(t)$ and, hence, Eq.~\eqref{EOM} is difficult to solve. In general, Eq.~\eqref{EOM} can be recast exactly or approximately into the form,
\begin{align}
	\frac{d}{dt}\tilde\rho_{\rm exc}(t)
	&=
	\sum_{\alpha}
	\mathrm{T}_+
	\int^t_0 ds 
	\tilde{A}_\alpha(t) e^{-b_\alpha(t-s)} \tilde{C}_\alpha(s)
	\tilde\rho_{\rm exc}(t),
\end{align}
where $\hat{A}_\alpha$ and $\hat{C}_\alpha$ are arbitrary operators and $b_\alpha$ is an arbitrary complex number, and do not necessarily have physical meaning. In this situation, Eq.~\eqref{EOM} is recast into the multidimensional recurrence form as \cite{Tanimura:1989bz,Tanimura:1990vh,Ishizaki:2005jw,Xu:2005ka,Xu:2007bm,Ishizaki:2009jg}
\begin{align}
	\frac{d}{dt}\hat\sigma(n_1,n_2,\dots)
	&=
	-(i\hat{\mathcal{L}}_{\rm exc}+n_1b_1+n_2b_2+\dots)\hat\sigma(n_1,n_2,\dots)
	\notag\\
	&+\hat{A}_1 \hat\sigma(n_1+1,n_2,\dots) + \hat{A}_2 \hat\sigma(n_1,n_2+1,\dots) +\dots
	\notag\\
	&+n_1 \hat{C}_1 \hat\sigma(n_1-1,n_2,\dots) + n_2\hat{C}_2 \hat\sigma(n_1,n_2-1,\dots) + \dots,
	\label{recurrence-eq}
\end{align}
where $\hat{\mathcal{L}}_{\rm exc}$ is the Liouvillian corresponding to the Hamiltonian, $\hat{H}_{\rm exc}$ in Eq.~\eqref{decomposed-hamiltonian}. In the above recurrence formula, only the $\hat\sigma(0,0,\dots)$ element is identical to the reduced density operator, $\hat\rho_{\rm exc}(t)$, while the other terms are auxiliary operators which do not necessarily have physical meaning. The recurrence in Eq.~\eqref{recurrence-eq} continues to infinity. However, the numerical calculations can converge at a finite stage. To terminate Eq.~\eqref{recurrence-eq} safely, we replace Eq.~\eqref{recurrence-eq} with
\begin{align}
	\frac{d}{dt}\hat\sigma(n_1,n_2,\dots)
	=
	-i\hat{\mathcal{L}}_{\rm exc}\hat\sigma(n_1,n_2,\dots),
\end{align}
for the integers $(n_1,n_2,\dots)$ satisfying $\sum_\alpha n_\alpha \gg \omega_c/\min_\alpha (b_\alpha)$, where $\omega_c$ is a characteristic frequency for $\hat{\mathcal{L}}_{\rm exc}$.\cite{Ishizaki:2005jw}

\section{Nonrephasing third-order response function\label{appendix-NR-response}}
\renewcommand{\theequation}{\ref{appendix-NR-response}.\arabic{equation}}

The nonrephasing 2D spectrum is calculated as
\begin{align}
	S_{\rm NR}(\omega_3, t_2, \omega_1)
	=
	\mathrm{Re}
	\int^\infty_0 dt_3 e^{i\omega_3t_3}
	\int^\infty_0 dt_1 e^{i\omega_1t_1}
	iR_{\rm NR}(t_3,t_2,t_1),
\end{align}
where the nonrephasing response function, $R_{\rm NR}(t_3,t_2,t_1)$, is expressed as\cite{Ishizaki:2006kf}
\begin{align}
	R_{\rm NR}(t_3,t_2,t_1)
	=
	\mathrm{Tr}
	\left\{
		\hat{\mu}_\leftarrow
		\hat{G}(t_3)
		\frac{i}{\hbar}
		\hat{\mu}_\rightarrow^\times
		\hat{G}(t_2)
		\frac{i}{\hbar}
		\hat{\mu}_\leftarrow^\times
		\hat{G}(t_1)
		\frac{i}{\hbar}
		\hat{\mu}_\leftarrow^\times
		\hat\rho_{\rm PPC}^{\rm eq}
	\right\}.
	\label{NR-response-function}
\end{align}
The operators, $\hat{\mu}_\rightarrow$ and $\hat{\mu}_\leftarrow$, are defined as
\begin{align}
	\hat{\mu}_\rightarrow 
	&= 
	\mu_1 \lvert 1 \rangle\langle g \rvert
	+
	\mu_2 \lvert 2 \rangle\langle g \rvert
	+
	\mu_1 \lvert 12 \rangle\langle 2 \rvert
	+
	\mu_2 \lvert 12 \rangle\langle 1 \rvert,
\\
	\hat{\mu}_\leftarrow 
	&= 
	\mu_1 \lvert g \rangle \langle 1 \rvert
	+
	\mu_2 \lvert g \rangle \langle 2 \rvert
	+
	\mu_1 \lvert 2 \rangle \langle 12 \rvert
	+
	\mu_2 \lvert 1 \rangle \langle 12 \rvert,
\end{align}
where $\mu_m$ is the transition dipole of the $m$-th pigment. We assume the Condon approximation and, thus, $\mu_m$ is independent of the environmental and nuclear DOFs. For simplicity, we set $\mu_m = 1$ and $\mu_1\cdot\mu_2=\cos\theta_{12}$, with $\theta_{12}$ being the angle between the two transition dipoles. Note that the directions of the subscript arrows in Eq.~\eqref{NR-response-function} correspond to those of the arrows in the double-sided Feynman diagrams depicting the nonrephasing response functions. Regarding the means of calculating the response function in Eq.~\eqref{NR-response-function} using the equation of motion approach given in Appendix~\ref{appendix-EOM}, we refer to Sec.~3 in Ref.~\onlinecite{Ishizaki:2006kf} and Sec.~5.5 in Ref.~\onlinecite{Tanimura:2006ga}.
We note that numerical results of the third-order response function in Eq.~\eqref{NR-response-function} do not depend on employed representations (e.g. the site representation or the energy eigenstate representation) because Eq.~\eqref{NR-response-function} is expressed as the trace of operators.

\section{Validity of the correlation function approach to describe vibronic excitons\label{validity-of-theory}}
\renewcommand{\theequation}{\ref{validity-of-theory}.\arabic{equation}}

Here, we comment on the validity of employing the correlation functions and the corresponding spectral density to describe the mixed electronic-vibrational state, instead of employing the vibronic exciton model.\cite{Womick:2011iw,Christensson:2012gp,Chenu:2013cf,Tiwari:2013dt} We consider the expression of F\"orster's EET rate\cite{Yang:2002ik,Jang:2004ic}
\begin{align}
	k^{\rm F}_{\rm A \gets D}
	=
	J_{\rm AD}^2\int^\infty_{-\infty}\frac{d\omega}{2\pi} 
	I_{\rm A}[\omega] E_{\rm D}[\omega],
	\label{forster-rate}
\end{align}
where $I_{\rm A}[\omega]$ and $E_{\rm D}[\omega]$ are the absorption line-shape of the energy acceptor molecule (A) and the emission line-shape of the donor (D), respectively. For simplicity, we employ the $J_{m,{\rm vib}}(\omega)$ given in Eq.~\eqref{spectral-density-delta} and assume the zero-temperature limit. In this situation, Eq.~\eqref{forster-rate} is recast into
\begin{align}
	k^{\rm F}_{\rm A \gets D}
	=
	\sum_{v =0}^\infty 
	\sum_{v'=0}^\infty
	\left(
		J_{\rm AD} 
		\langle \chi_{ev }^{\rm A} \vert \chi_{g0}^{\rm A} \rangle
		\langle \chi_{gv'}^{\rm D} \vert \chi_{e0}^{\rm D} \rangle
	\right)^2
	\int^\infty_{-\infty}
	\frac{d\omega}{2\pi} 
	I_{\rm A}^{(v)}[\omega] E_{\rm D}^{(v')}[\omega],
	\label{Forster-decomp}
\end{align}
where
\begin{gather}
	\lvert \langle \chi_{a' v}^{m} \vert \chi_{a 0}^{m} \rangle \rvert^2
	=
	\frac{S^v}{v!}e^{-S},
\\
	I_{\rm A}^{(v)}[\omega]
	=
	\pi\delta(\omega - (\Omega_{\rm A}-S \omega_{\rm vib}+v\omega_{\rm vib})),
	\label{absorption-decomp}
\\
	E_{\rm D}^{(v')}[\omega]
	=
	\pi\delta(\omega - (\Omega_{\rm D}-S \omega_{\rm vib}-v'\omega_{\rm vib})).
	\label{emission-decomp}
\end{gather}
In the above, $\lvert \langle \chi_{a' v'}^{m} \vert \chi_{a v}^{m} \rangle \rvert^2$ is the Franck-Condon factor associated with the vibronic transition from the $v$-th vibrational level of the $a$-th electronic state, $\lvert \chi_{av}^m \rangle$, to the $v'$-th vibrational level of the $a'$-th electronic state, $\lvert \chi_{a'v'}^m \rangle$, of the $m$-th pigment. Equation~\eqref{Forster-decomp} indicates that, in the case of $\Omega_{\rm A} + v\omega_{\rm vib} = \Omega_{\rm D} - v' \omega_{\rm vib}$, the two vibronic transitions, $\lvert \varphi_{{\rm D}e} \rangle \lvert \chi_{e0}^{\rm D} \rangle \to \lvert \varphi_{{\rm D}g} \rangle \lvert \chi_{gv'}^{\rm D} \rangle$ and $\lvert \varphi_{{\rm A}g} \rangle \lvert \chi_{g0}^{\rm A} \rangle \to \lvert \varphi_{{\rm A}e} \rangle \lvert \chi_{ev}^{\rm A} \rangle$, resonate, and excitation transfer then occurs between them with the coupling constant
\begin{align}
		J_{\rm AD} 
		\langle \chi_{ev }^{\rm A} \vert \chi_{g0}^{\rm A} \rangle
		\langle \chi_{gv'}^{\rm D} \vert \chi_{e0}^{\rm D} \rangle.
\end{align}
This is consistent with the approach using the vibronic exciton Hamiltonian model, \cite{Womick:2011iw,Renger:2011hj,Christensson:2012gp,Chenu:2013cf,Tiwari:2013dt} and, thus, the present correlation function approach certainly describes the interaction between the vibronic transitions. Furthermore, it should be emphasized that the current approach is capable of describing vibrational relaxations, vibrational states in both the electronic ground and excited states, and the influence of the environmental dynamics in a natural and consistent manner.


\newpage

\end{document}